\documentstyle[epsfig,12pt]{article}
\def\aleq{\; \raisebox{.5ex}{$<$}\hspace{-3.2mm}\raisebox{-.5ex}{$\sim$}\; }
\def\ageq{\; \raisebox{.5ex}{$>$}\hspace{-3.2mm}\raisebox{-.5ex}{$\sim$}\; }
\def\sD{{\cal D}\hspace{-2.5mm}/}

\begin{document}

\begin{flushright}
OCHA-PP-95
\end{flushright}
\vspace{.2cm}
\begin{center}
{\Large\bf Chaotic inflation with running nonminimal coupling }
\end{center}

\vspace{.3cm}
\begin{center}
{Toshifumi Futamase\\
{\small\sl Astronomical Institute, Tohoku University, 
Sendai, 980-77  Japan}\\
Masahiro Tanaka\\ 
{\small\sl Department of Physics, Ochanomizu University, Tokyo, 112  Japan}}
\end{center}

\vspace{.5cm}

\begin{abstract}
We have found a successful model of chaotic inflation with an inflaton 
coupled nonminimally with gravity. 
The nonminimal coupling constant $\xi$ runs with the evolution of 
the inflaton. The running nature of the coupling leads naturally to the 
situations where the coupling becomes small enough 
to have sufficient period of 
the inflation to resolve the cosmological puzzles.
\end{abstract}


\section{Introduction}
Chaotic inflation\cite{l,l1} 
is known to be a successful scenario of inflation. Although  
its setup is regarded as simpler than other scenarios, 
it still needs fine-tuning of some parameters as others do. 
The major two are the inflaton self-coupling constant\cite{l1}, 
and the nonminimal coupling constant to gravity\cite{f&m,f&r&m,f,f&u}. 

As far as the former is concerned, we may identify the direction 
which contains only quadratic terms, e.g. mass term, 
with inflaton and then we can avoid the fine-tuning\cite{m&s&y&y}.  
The direction is called flat direction,\footnote{
We mean the direction without interaction terms by
  flat direction. We assume that non-renormalizable terms are also
  vanishing because the inflaton will be the scalar field with the most
  flat potential.} 
which frequently appears in supersymmetric (SUSY) models. Interactions still
exist in the theory but such terms do not appear along flat direction.

Unfortunately so far no mechanism is known to avoid the latter fine-tuning.  
Namely we have the freedom to specify how the inflaton 
couples with gravity, namely how to choose the nonminimal coupling constant. 
The nonminimal coupling with gravity leads to a soft SUSY
breaking\cite{t2} and is needed for renormalization.
In the case of positive nonminimal 
coupling constant it has been shown that 
the chaotic inflationary scenario does not work unless it 
is sufficiently small\cite{f&m,f&r&m,f}. On the other hand, 
the scenario does work in the case 
of negative coupling constant. In
particular Fakir and Unruh\cite{f&u} show that a model with a large negative
nonminimal coupling constant gives the appropriate amplitude of
density perturbation without making any fine-tuning for the self-coupling.

However, the previous studies on the chaotic inflation scenario with
nonminimal coupling are restricted to the classical teratment. 
 The aim of this paper is to 
take into account of the quantum effect on the coupling 
which makes the coupling constant running with the inflaton field.  
The running nature of the coupling constant may allow us to have new  
possibility which is not available in classical treatments. 
Namely, there are always some regions in spacetime where the coupling 
constant becomes small enough to have 
sufficient period of inflation. 
We shall study this possibility in SUSY minded models and find that 
this is the case with reasonable range of parameters.


\section{Chaotic inflation with nonminimal coupling to gravity}

According to the above argument, we shall take the following 
action for our model of the nonminimally coupled inflaton 
in Einstein gravity\cite{a&l&o,b&c&c&m&n}:
\begin{eqnarray}
S=-\frac{1}{2}\int d^4x\sqrt{-g}\biggl[ 
\frac{{M_{Pl}}^2}{8\pi}{\cal R}+\partial _{\mu }\phi \cdot \partial ^{\mu}\phi 
+m^2\phi ^2-\xi {\cal R}\phi ^2\biggr] ,
\label{1}
\end{eqnarray}
where  $M_{Pl}=1/\sqrt{G}$ and $\xi $ is the nonminimal coupling constant. 
Conformal invariance yields $\xi =1/6$. 
The above action is considered to be the bosonic sector of a Wess-Zumino
model in curved spacetime and the Einstein term. The nonminimal
coupling is a soft SUSY breaking term as mass term.
The equations of motion are
\begin{eqnarray}
& &(\Box -m^2+\xi {\cal R})\phi =0
\label{9}\\
& &{\cal R}_{\mu \nu}-\frac{1}{2}{\cal R}g_{\mu \nu}
=\frac{8\pi}{M_{Pl}^2}{\cal T}^{eff}_{\mu \nu},
\label{10}
\end{eqnarray} 
where 
\begin{eqnarray}
{\cal T}^{eff}_{\mu \nu}&=&\frac{{M_{Pl}}^2}{{M_{Pl}}^2 -8\pi \xi \phi ^2}
\biggl[ (1-2\xi)\partial _{\mu }\phi \cdot \partial _{\nu}\phi 
+\biggl( 2\xi -\frac{1}{2}\biggr) 
\partial _{\rho}\phi \cdot \partial ^{\rho}\phi 
g_{\mu \nu}\nonumber\\
& &-2\xi \phi \nabla _{\mu}\partial _{\nu}\phi 
-\frac{1}{2}m^2\phi ^2g_{\mu \nu} -2\xi g_{\mu \nu}\partial ^{\rho}\phi 
\cdot \partial _{\rho}\phi \biggr] .
\label{11}
\end{eqnarray} 
Since we focus on small regions in the spacetime in which the inflaton 
distributes homogeneously and examine if such a region will undergo 
sufficent period of the inflation, we shall assume that the region is 
homogenous and isotropic, and thus can be described by 
Robertson-Walker metric. Then we have the Friedmann and Raychaudhuri
equation for such a region:
\begin{eqnarray}
& &H^2+\frac{k}{a^2}=\frac{\dot{\phi} ^2+12\xi H\dot{\phi }\phi +m^2\phi ^2 }
{6({M_{Pl}}^2/8\pi -\xi \phi ^2)},
\label{12}\\
& &\dot{H}+H^2=\frac{(6\xi -2)\dot{\phi} ^2+m^2\phi ^2 
+6\xi \ddot{\phi}\phi +6\xi H\dot{\phi}\phi }
{6({M_{Pl}}^2/8\pi -\xi \phi ^2)} .
\label{13}
\end{eqnarray}
where $H = {\dot a}/a $ is the Hubble parameter of the region, 
with $a$ the scale factor.
Inserting the Eqs.~$(\ref{12})$ and$(\ref{13})$ into Eq.~$(\ref{9})$, 
we obtain the Klein-Gordon type equation for the inflaton:

\begin{eqnarray}
& &\ddot{\phi }+3H\dot{\phi }+\Biggl[ m^2 
+\frac{ ( 6\xi -1) \dot{\phi} ^2
+2m^2\phi ^2 +6\xi (\ddot{\phi } +3H\dot{\phi })\phi }
{{M_{Pl}}^2/8\pi -\xi \phi ^2}\xi \Biggr] \phi =0.
\label{14}
\end{eqnarray}

For a negative $\xi $ 
it is known that there are the two saddle points in the phase
space, where $\phi_{cr} = \pm M_{Pl}/\sqrt{8\pi | \xi | }$ and 
$\dot{\phi }=0$\cite{a&l&o,b&c&c&m&n}.
The origin is the unique stable point in the phase space. If the
inflaton 
initially satisfy the condition 
\begin{eqnarray}
-\frac{M_{Pl}}{\sqrt{8\pi | \xi |}} < \phi < 
\frac{M_{Pl}}{\sqrt{8\pi | \xi |}},
\label{23.1}
\end{eqnarray}
 chaotic inflation may occur. 
Otherwise, the scalar field keeps growing exponentially: inflation
occurs but never terminates\cite{a&l&o}.

On the other hand, for a positive $\xi $ the anisotropic shear 
diverges as the inflaton approaches at the following points 
without bound{\cite{f&r&m,s} 
\begin{eqnarray}
\phi_{cr} =\pm \frac{M_{Pl}}{\sqrt{8\pi \xi }}.
\label{23.2}
\end{eqnarray}
The evolution of the region with $\phi$ larger than $\phi_{cr}$ 
will terminate at $\phi_{cr}$, and such regions do not evolve into 
pur present Universe. 
We shall call this points as the critical points.
Thus we shall only pay attention to the region with $\phi$ lying between 
the two critical points\cite{f&m,f&r&m,f}. 
This gives the same constraint as the case with negative $\xi$.

It has been known that the minimal initial value of the inflaton 
is about $ 5M_{Pl}$ to realize sufficient period of inflation in the 
framework of chaotic inflationary scenario.  
Then the above condition  $(\ref{23.1})$ and $(\ref{23.2})$ 
with  $|\phi | \sim 5M_{Pl}$ 
give us the condition  $| \xi |\simeq 10^{-3}$ for successiful 
chaotic inflationary senario with nonminimal coupling. In
Ref.~\cite{f&m} they demand also more severe constraint for natural 
realization of inflation, because they consider that inflaton probably
have Planck energy density initially. We do not adopt such a view,
instead we assume that the initial
value is distributed randamly below Planck energy density. Therefore 
we just need $|\phi | \ageq 5M_{Pl}$ for the initial condition.


\section{Running nonminimal coupling }

We adopt the one-loop effective action along a flat direction 
of the Wess-Zumino 
model\cite{s&t} as the Lagragian (\ref{1}).
The superpotential is $W=g\Phi _1\Phi _2\Phi _2/2$.
 The action is written as
\begin{eqnarray}
e^{-1}{\cal L}&=&-(\partial _{\mu}\phi _i)^*(\partial ^{\mu}\phi _i)
-V(\phi _1,
\phi _2) \nonumber \\
& &+\frac{1}{2}(\bar{\psi _1}\; \bar{\psi _2})
\bigl[ i\sD +M(\phi _1,\phi _2)\bigr] \left( \begin{array}{c}\psi _1\\ \psi _2
 \end{array}  \right) ,\label{3,1}
\end{eqnarray}   
where ${\cal D}_{\mu} $ is 
covariant derivative and the potential term $V$ is defined 
as 
\begin{eqnarray}
V(\phi _1,\phi _2)&=&
(m_1^2-\xi _1{\cal R})\phi _1^*\phi _1+\frac{1}{4}|g^2|(\phi _2^*
\phi _2)^2 \nonumber \\
& &+(m_2^2-\xi _2{\cal R})\phi _2^*\phi _2
+|g^2|(\phi _1^* \phi _1)(\phi _2^*\phi _2),
\label{3,2}
\end{eqnarray}
and the mass matrix $M(\phi _1,\phi _2)$ is given by 
\begin{eqnarray}
M(\phi _1,\phi _2)=\left( \begin{array}{cc}
0 & {\rm Re}(g\phi _2)-{\rm Im}(g\phi _2)\gamma ^5 \\
{\rm Re}(g\phi _2)-{\rm Im}(g\phi _2)\gamma ^5 & 
{\rm Re}(g\phi _1)-{\rm Im}(g\phi _1)\gamma ^5
\end{array} \right) .\label{3,3}
\end{eqnarray}
${\cal R}$ is the scalar curvature and ${\cal R}=12H^2$ 
in de Sitter spacetime .
 In the scalar potential $(\ref{3,1})$, we have introduced the soft 
SUSY breaking mass terms $m_i^2\phi _i^*\phi _i$ and the nonminimal curvature 
couplings $\xi _i{\cal R}\phi _i^*\phi _i$,
in addition to the minimal extension of 
the Wess-Zumino model in curved spaces. Since the nonminimal coupling 
receives the renormalization as will be seen below, the bare term 
$\xi_i{\cal R}\phi _i^*\phi _i $ is necessary for this model to be 
renormalizable. 
From the tree potential $(\ref{3,2})$, we see $\phi _2 =0$ is actually a flat 
direction in this model; namely the potential energy remains flat for any 
values of $\phi _1$ except for the SUSY breaking mass term and the 
nonminimal coupling term.

Now let us consider the one-loop effective potential. 
We decompose the scalar fields as 
\begin{eqnarray}
\phi _1=\frac{\phi _1}{\sqrt{2}}+\varphi _1+i\varphi _2,\; \; \phi _2=
\frac{\phi _2}{\sqrt{2}}+\varphi _3+i\varphi _4,
\label{3,4}
\end{eqnarray}
where all the fields are real and $\phi _i$ are the classical fields. 

Using the DeWitt-Schwinger technique\cite{b&d} we obtain 
\begin{eqnarray}
{\cal V}_{eff}(\phi _1,\phi _2)&=&
\frac{1}{2}(m_1^2-\xi _1{\cal R}){\phi _1}^2+
\frac{1}{2}(m_2^2-\xi _2{\cal R}){\phi _2}^2
+\frac{1}{16}g^2{\phi _2}^4 
+\frac{1}{4}g^2{\phi _1}^2{\phi _2}^2 \nonumber\\
& &+\frac{g^2}{32\pi ^2}
\ln \frac{{\phi _1}^2+\cdots }{\Lambda ^2}\Biggl[ 
\biggl\{ {m_2}^2+\biggl( \xi _2-\frac{1}{4}\biggr) \biggr\} 
({\phi _1}^2+{\phi _2}^2)\nonumber\\
& &+\frac{1}{4}g^2{\phi _1}^2{\phi _2}^2
+\frac{1}{16}g^2{\phi _2}^4\Biggr] +\frac{g^2}{32\pi ^2}
\ln \frac{{\phi _2}^2+\cdots }{\Lambda ^2}\Biggl[
\biggl\{ {m_1}^2\nonumber\\
& &+\biggl( \xi _1-\frac{1}{4}\biggr) \biggr\} 
{\phi _2}^2+\frac{1}{4}g^2{\phi _1}^2{\phi _2}^2\Biggr] 
\end{eqnarray}
in the flat limit, where we have omitted the $\phi _i$-independent 
convergent terms which are irrelevant in our present analysis. 
 $(\cdots )$ do not contain $\phi _1$-dependent term. 
Likewise\cite{b,t3}, the renormalized wave functions are 
defined as
\begin{eqnarray}
& &{\phi _{1R}}^2= \biggl( 1+\frac{g^2}{32\pi ^2}\ln \frac{\Lambda
  ^2}{{\phi _1}^2+\cdots}\biggr) {\phi _1}^2,
\label{6.1}\\
& &{\phi _{2R}}^2= \biggl( 1+\frac{g^2}{32\pi ^2}\ln \frac{\Lambda
  ^2}{{\phi _1}^2+\cdots}+\frac{g^2}{32\pi ^2}\ln \frac{\Lambda
  ^2}{{\phi _2}^2+\cdots}\biggr) {\phi _2}^2.
\label{6.2}
\end{eqnarray}
Then we define the renormalized parameters
\begin{eqnarray}
& &{g_R}^2=\biggl( 1
+\frac{3g^2}{32\pi ^2}\ln \frac{{\phi _1}^2+\cdots }{\Lambda ^2}\biggr)
g^2,
\label{6.3}\\
& &{m_{1R}}^2={m_1}^2+
\frac{g^2}{32\pi ^2}\biggl( \ln \frac{{\phi _1}^2+\cdots }{\Lambda ^2}\biggr)
({m_1}^2+2{m_2}^2),
\label{6.4}\\
& &{m_{2R}}^2={m_2}^2+
\frac{3g^2}{32\pi ^2}\biggl(\ln \frac{{\phi _1}^2+\cdots }{\Lambda
  ^2}\biggr) {m_2}^2,
\label{6.5}\\
& &\xi _{1R}=\xi _1+\frac{g^2}{32\pi ^2}\biggl( \ln \frac{{\phi _1}^2+\cdots }
{\Lambda ^2}\biggr) \biggl( \xi _1+2\xi _2-\frac{1}{2}\biggr) , 
\label{6.6}\\
& &\xi _{2R}=\xi _2+\frac{g^2}{32\pi ^2}\biggl( \ln \frac{{\phi _1}^2+\cdots }
{\Lambda ^2}\biggr) \biggl( 3\xi _2-\frac{1}{2}\biggr) ,
\label{6.7}
\end{eqnarray}
where we present only correction with $\phi _1$ (the flat direction).
The renormalizaiton groups are
\begin{eqnarray}
& &\phi \frac{d\, g^2}{d\phi}=\frac{3g^4}{16\pi ^2},
\label{6.8}\\
& &\phi \frac{d\, {m_1}^2}{d\phi}=\frac{g^2}{16\pi ^2}({m_1}^2+2{m_2}^2),
\label{6.81}\\
& &\phi \frac{d\, {m_2}^2}{d\phi}=\frac{3g^2}{16\pi ^2}{m_2}^2,
\label{6.82}\\
& &\phi \frac{d\, \xi _1}{d\phi}=\frac{g^2}{16\pi ^2}\biggl( \xi _1+
2\xi _2-\frac{1}{2}\biggr),
\label{6.83}\\
& &\phi \frac{d\, \xi _2}{d\phi}=\frac{3g^2}{16\pi ^2}\biggl( \xi _2
-\frac{1}{6}\biggr) ,
\label{6.84}
\end{eqnarray}
where we simply denote $\phi _1$ by $\phi $ and hereafter all the
quantities are renormalized.
The renormalization group analysis of the nonminimal coupling says
\begin{eqnarray}
\xi _2(\phi )
&=&\frac{g^2(\phi)}{g^2(M)}\biggl[ \xi _2(M)-\frac{1}{6}\biggr]
+\frac{1}{6}.
\label{7.1}
\end{eqnarray}
where 
\begin{eqnarray}
g^2(\phi) &=&g^2(M)\biggl[ 1-\frac{3g^2(M)}{32\pi ^2}\ln \frac{\phi
  ^2}{M^2}\biggr] ^{-1} .
\label{7.2}
\end{eqnarray}
Eqs.~(\ref{6.83}) and (\ref{6.84}) allow us to assume 
$\xi _1=\xi _2\equiv \xi $. 
This theory is free and conformal invariant in the infrared
limit; $ \phi \rightarrow 0$.\footnote{The running of the mass parameter
  is ${m_2}^2(\phi )=
\{ g^2(\phi )/g^2(M)\} {m_2}^2(M)$. If we choose a right-handed 
sneutrino\cite{m&s&y&y} as the inflaton, it is plausible to assume 
$m(M_{Pl})\simeq 10^{13}GeV$. $m(\phi )$ does not change its order of
magnitude drastically in the relevant scale.}  
In other words, in this limit
 the free scalar field propagates along the light cone.\cite{f} 
The form of Eq.~($\ref{7.1}$) is universal no matter 
if its conformal invariance appears in the ultraviolet or infrared 
region.\cite{b&o,b&o&s} \footnote{What we actually do is called
  renormalization group improvement of effective potential.
\cite{b&o&s,c&w,b&o2}}

As far as $g^2(\phi )<1$ and $3g^2(M)\ln(\phi ^2/M^2)/32\pi^2 <1$, 
one-loop calculation is reliable.

Contrary to $m$, identification of the inflaton with some elementary 
particle cannot constrain $\xi$.
If we define the renormalization point $M$ as
\begin{eqnarray}
\xi (M) =0,
\label{7.4}
\end{eqnarray}
 
then Eq.~(\ref{7.2}) becomes
\begin{eqnarray}
\xi (\phi )=\frac{1}{6}-
\frac{1}{6}\biggl[ 1-\frac{3g^2(M)}{32\pi ^2}\ln \frac{\phi
  ^2}{M^2}\biggr] ^{-1}.
\label{7.41}
\end{eqnarray}
We use the Eq.~(\ref{7.41})
\footnote{In other words we choose the flow which decreases 
monotonically to the negative infinity as 
$3g^2(M)\ln (\phi ^2/M^2)/32\pi ^2 \to 1$.} for the later argument,
where we assume Eq.~(\ref{7.41}) is applicable up to the Planck energy
density: $\phi \simeq M_{Pl}^2/m$.


\section{Condition on Yukawa coupling}
In this section we clarify the condition on the Yukawa coupling constant 
for the realization of a 
successiful inflationary scenario using the above results.
First of all we assume that the inflaton must have the initial 
value\cite{l,l1,f&m}:
\footnote{
Hereafter we concentrate on the positive side of $\phi $.}
\begin{eqnarray}
\phi \ageq 5M_{Pl}
\label{7.32}
\end{eqnarray}
As disccused in section 2 this is needed for a sufficient period of inflation.
It is convenient to devide the range of $M$ as 

$5M_{Pl}<M$ and $M\aleq 5M_{Pl}$, 
which we call them case (I) and case (II), respectively.

\subsection{case (I)}

In this case $\xi (5M_{Pl}) $ is positive. Then during evolution of
inflaton $\xi(\phi )$ is always positive.
Since we must avoid the critical point, we obtain the condition:
\begin{eqnarray}
\frac{M_{Pl}}{\sqrt{8\pi \xi(\phi )}}> \phi
\end{eqnarray}
during the whole evolution of $\phi $, $\phi \aleq 5M_{Pl}$. 

The above condition is rewritten as ( See Figure~\ref{f2})
\begin{eqnarray}
\frac{M_{Pl}}{2}\sqrt{\frac{3}{\pi}-\frac{32\pi }
{g^2(M)\ln {(\phi /M)^2}}} >\phi 
\label{7.320}
\end{eqnarray}
for all $\phi \aleq 5M_{Pl}$.

Numerical results say that
 for $10\aleq M/5M_{Pl}<10^2$ we need $g^2(M)\aleq 1$ and 
for $10^2\aleq M/5M_{Pl}\aleq 10^6$ we need $g^2(M)\aleq 10^{-1}$. We
will not consider the range $M/5M_{Pl}> 10^6$ because the energy density 
becomes larger than Planck energy density for such ranges.

\begin{figure}[h]
\begin{center}
\epsfig{file=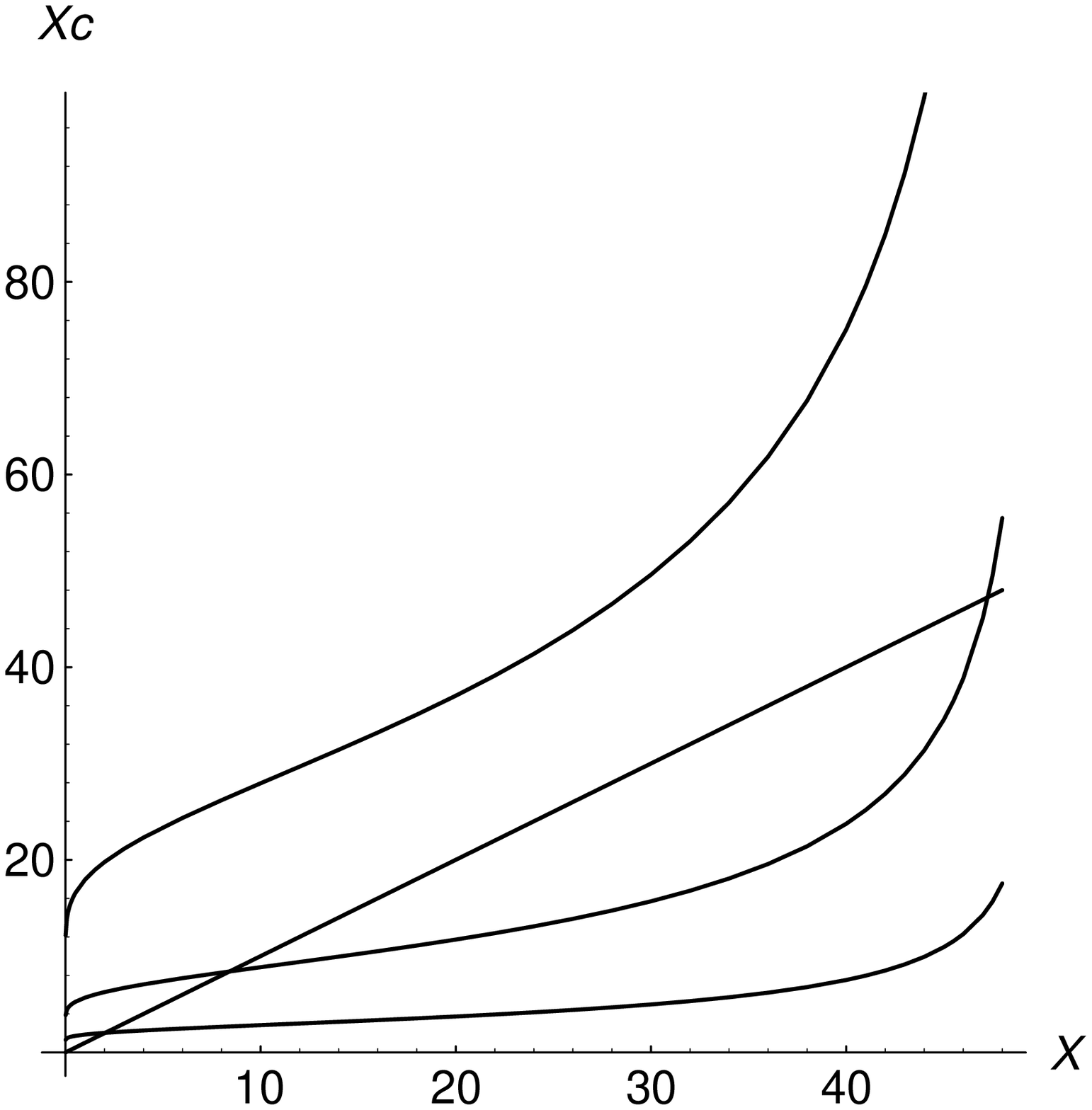,width=12cm}
\end{center}
\caption{$x=\phi /M_{Pl}$, $x_c=\phi _{cr}/M_{Pl}$, and 
$M/M_{Pl}=50$. 
The upper line corresponds to the case $g^2(M)=10^{-2}$,
 the middle line to the case $g^2(M)=10^{-1}$, and the lower to the 
case $g^2(M)=1$. The straight line indicates $x_c=x$: $\phi _{cr}=\phi$. 
The former two lead to a successful inflation because 
$\phi _{cr}>\phi $ for $\phi < 5M_{Pl}$.}
\label{f2}
\end{figure}

\subsection{case (II)} 

In this case $\xi(5M_{Pl})$ is negative.
$5M_{Pl}$ should be smaller than the unstable point and during 
the subsequent evolution $\phi $ must avoid the singularity: 
\begin{eqnarray}
\frac{M_{Pl}}{\sqrt{8\pi |\xi(5M_{Pl})|}}&>& 5M_{Pl}
\label{7.321}\\
\frac{M_{Pl}}{\sqrt{8\pi \xi(\phi )}}&>& \phi
\label{7.322}
\end{eqnarray}
for all $\phi <M$. Eq.~(\ref{7.321}) is sufficient for $\phi >M$ since
the value of the unstable point ($M_{Pl}/\sqrt{8\pi |\xi(\phi )|} $)
monotonially decreses in such a region.
 
Eq.~(\ref{7.321}) gives the condition on $\xi$ as
\begin{eqnarray}
|\xi (5M_{Pl})|\aleq 10^{-3},
\label{7.33}
\end{eqnarray}
which is the same as the condition for the constant $\xi$\cite{f&m}.
Roughly this gives us the following condition on $g(M)$ (See Figure~\ref{f3}) 
\begin{eqnarray}
g^2(M)\aleq \frac{1}{\ln(5M_{Pl}/M)^2}
\label{7.34}.
\end{eqnarray} 
This gives us the condition on $g(M)$.
For example, $g^2(M) \sim 10^{-1}$ for $M/5M_{Pl} \sim 10^{-2}$, 
$g^2(M) \sim 10^{-2}$ for 
$M/5M_{Pl}\sim 10^{-22}$ and so on. We can see the condition 
is not unreasonable at all in a SUSY model\cite{m&s&y&y}. 

The second condition (\ref{7.322})
is automatically satisfied if perturbation is reliable: $g(\phi )<1$(See 
Figure~\ref{f3}).

\begin{figure}[h]
\begin{center}
\epsfig{file=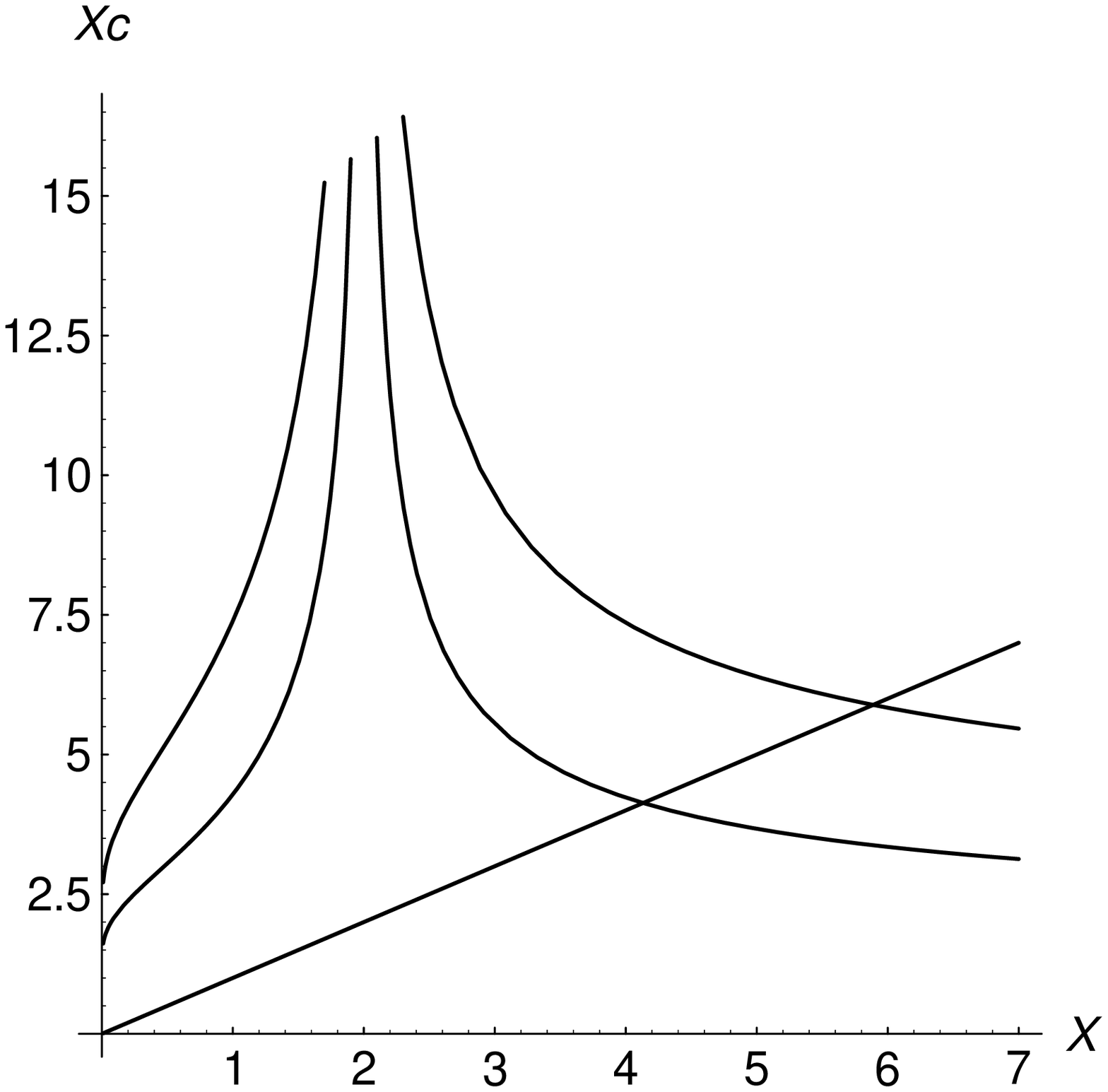,width=11cm}
\end{center}
\caption{ $x=\phi /M_{Pl}$, $x_c=\phi _{cr}/M_{Pl}$, and 
$M/M_{Pl}=2$. 
The upper line corresponds to the case $g^2(M)=1/3$ and the lower does to the 
case $g^2(M)=1$. The straight line indicates $x_c=x$: $\phi _{cr}=\phi$. 
The former leads to a successful inflation because 
$\phi _{cr}>\phi $ for $\phi < 5M_{Pl}$. From 
Eq.(\ref{7.34}) we
can find with this value of $ M $ we need 
$g^2(M)< 0.54$ approximately.
}
\label{f3}
\end{figure}


\section{Conclusions and Discussion}
We have found that chaotic inflation by nonminimal coupled inflaton
is naturally realized by taking into account of the running nature 
of the coupling constant. The condition to have a successful inflation is 
just $g(M)\aleq 10^{-1}$ around a wide range of $M$ where $\xi
(M)=0$ which is not unreasonable in some models. It also implies that efficient
reheating may be possible. Nonminimal coupling constant becomes to be 
small enough
($|\xi|\aleq 10^{-3}$) 
when inflation starts and evolves to the conformal value, where the
universe is radiation dominant.

The above argument is true to all the models which have flat directions 
since the form of the renormalization group equation is universal.

A model with $-\xi \ageq 10^4$ is considered to circumvent fine-tuning
of the self-coupling constant in $\lambda \phi ^4$ theory
\cite{f&u}.
But such a large $| \xi|$ is realized by a large $\phi $ which causes very
high energy density larger than the Planck energy
density. Accordingly, such a model seems to be not so feasible. However, 
the more detailed investigation is neccessary to draw any definte conclusion 
for such cases.

\section*{Acknowledgement}
The authors thank to M. Hotta and K. Kumekawa for valuable comments.

\appendix

\end{document}